\documentclass[aps,prd,reprint,superscriptaddress,nofootinbib,floatfix]{revtex4-2}

\usepackage{graphicx}
\usepackage{dcolumn}
\usepackage{bm}

\usepackage{amsmath}
\usepackage{amssymb}
\usepackage{mathrsfs}
\usepackage{braket}
\usepackage{slashed}
\usepackage[makeroom]{cancel}
\usepackage{dsfont}

\usepackage{xcolor}

\usepackage{balance}

\def\fun#1#2{\lower3.6pt\vbox{\baselineskip0pt\lineskip.9pt
  \ialign{$\mathsurround=0pt#1\hfil##\hfil$\crcr#2\crcr\sim\crcr}}}

\def\lsim{\mathrel{\rlap{\raise 2.5pt \hbox{$<$}}\lower 2.5pt\hbox{$\sim$}}}
\def\gsim{\mathrel{\rlap{\raise 2.5pt \hbox{$>$}}\lower 2.5pt\hbox{$\sim$}}}

\newcommand{\comment}[1]{}

\begin{document}

\title{Local Minimum of Spin-Sector Magic at the CP-Conserving Point in Low-Energy Neutron-Proton Scattering}

\author{Cihang Li}
\affiliation{School of Physics, Peking University, Beijing, 100871, China}

\author{Teng Ma}
\affiliation{International Center for Theoretical Physics Asia-Pacific (ICTP-AP), University of Chinese Academy of Sciences, 100190 Beijing, China}
\affiliation{Taiji Laboratory for Gravitational Wave Universe, University of Chinese Academy of Sciences (UCAS), Beijing, 100190, China}

\author{Mingdi Zhu}
\affiliation{International Center for Theoretical Physics Asia-Pacific (ICTP-AP), University of Chinese Academy of Sciences, 100190 Beijing, China}
\affiliation{Taiji Laboratory for Gravitational Wave Universe, University of Chinese Academy of Sciences (UCAS), Beijing, 100190, China}

\date{\today}

\begin{abstract}
We study Magic generation in elastic neutron-proton scattering within a leading low-energy spin-sector ansatz that retains the one-pion-exchange spin structures and treats each scattering direction as a conditional two-qubit spin map. We show that the direction-averaged Magic is locally minimized at the CP-conserving (CPC) point $\bar\theta=0$ at the Clifford point $f_{\rm CPC}=\pi/4$, and for the representative non-Clifford CPC backgrounds analyzed here. At $f_{\rm CPC}=\pi/4$, the CPC spin map reduces to SWAP up to a phase and therefore generates zero Magic from stabilizer inputs. We further evaluate the complete spin-sector Magic functional by averaging over all 60 two-qubit stabilizer inputs and over scattering directions, and find that the curvature at $\bar\theta=0$ is positive only within specific windows of the effective CPC phase $f_{\rm CPC}$. These results identify the CPC point as a local Magic minimum within the restricted low-energy spin sector considered here.

\end{abstract}

\maketitle

\section{Introduction}
Quantum-information diagnostics provide a complementary way to characterize scattering dynamics. Entanglement alone does not measure classical intractability: by the Gottesman-Knill theorem, Clifford circuits acting on stabilizer states can generate substantial entanglement while remaining efficiently classically simulable~\cite{gottesman1998heisenbergrepresentationquantumcomputers,Nielsen_Chuang_2010}. The relevant nonclassical resource is nonstabilizerness, or Magic, which quantifies non-Clifford computational complexity and is required for universal quantum computation~\cite{Bravyi_2005}. This makes Magic a natural observable for asking how fundamental interactions generate genuinely quantum computational structure, beyond the correlations already captured by entanglement-based measures.

Recent work has begun to apply quantum-information diagnostics to high-energy processes, treating scattering not only as dynamics but also as a resource-generating quantum channel~\cite{Low_2021,LIU2024138899,carena2023entanglementsuppressionenhancedsymmetry,Chang_2024,Kowalska_2024,mcginnis2025symmetryentanglementsmatrix,liu2025parameterinferencefinalstateentanglement,Busoni_2025,Liu:2025qfl,Liu:2025bgw}. Previous studies of entanglement already linked emergent symmetries to extrema of quantum correlations~\cite{LIU2024138899,Low_2021,carena2023entanglementsuppressionenhancedsymmetry,Carena:2025wyh,Chang_2024,Kowalska_2024,mcginnis2025symmetryentanglementsmatrix,liu2025parameterinferencefinalstateentanglement}. More recently, Magic-based analyses have suggested that physically realized Standard Model parameters can lie near minima of non-Clifford resource generation~\cite{Busoni_2025,Liu:2025qfl,Liu:2025bgw}. In Quantum Electrodynamics, electron-muon scattering processes generate comparatively little Magic~\cite{Liu:2025qfl}. In the electroweak sector, the experimentally measured weak mixing angle at $\sqrt{s}=m_Z$ was found to lie near a local Magic minimum in Møller scattering~\cite{Liu:2025bgw}. These examples motivate a sharper question for strong interactions: can the CP-conserving point of QCD also be distinguished by an extremum of non-Clifford resource generation?

Motivated by this pattern, we ask whether an analogous information-theoretic extremum appears in a QCD-motivated setting connected to one of the most enduring puzzles in particle physics: the Strong CP Problem. The QCD Lagrangian admits a CP-violating (CPV) topological term parameterized by the physical vacuum angle $\bar{\theta}$~\cite{tHooft:1976rip,Callan:1976je}. While theoretical considerations allow $\bar{\theta}$ to take any value, experimental constraints from the neutron electric dipole moment (nEDM) impose an unnatural bound of $|\bar{\theta}| \lesssim 10^{-10}$~\cite{Abel_2020}. In the absence of a protective symmetry, this fine-tuning remains unexplained within the Standard Model, motivating dynamical solutions such as the Peccei-Quinn axion mechanism~\cite{Peccei:1977hh,Wilczek:1977pj,PhysRevLett.40.223,Peccei_2008}. This longstanding puzzle continues to motivate both axion searches and complementary theoretical viewpoints on why the CPC point may be dynamically or structurally distinguished~\cite{Peccei_2008}. Low-energy neutron-proton scattering provides a useful testing ground for this question because its spin dependence is well organized in terms of singlet-triplet structures and controlled CPV pion-nucleon operators.

In this work, we address that question in a controlled but restricted setting. We analyze low-energy neutron-proton ($np$) scattering within a chiral spin-sector ansatz, treat the spin-dependent interaction as a conditional two-qubit map, and quantify its non-Clifford content using the Stabilizer Rényi Entropy $M_2$~\cite{Leone_2022}. Our strategy has two layers. First, we compute the direction-averaged Magic generated by the effective spin map and show numerically that, in the perturbative regime of small $\bar{\theta}$, the averaged Magic has a stable local minimum at the CPC point $\bar{\theta}=0$ for the Clifford point and for representative non-Clifford CPC backgrounds. Second, we evaluate the finite spin-sector Magic functional of this framework by averaging over all two-qubit stabilizer inputs and over scattering directions, which lets us determine explicitly the curvature windows in the effective CPC phase for which that local minimum persists. Throughout, the claim is restricted to the OPE-retained low-energy spin sector rather than to the full chiral EFT or QCD scattering matrix.

\section{Magic and Stabilizer Rényi Entropy} \label{Sec:magic}
To define the non-Clifford resource measured below, we use the stabilizer formalism~\cite{PhysRevLett.109.230503,Veitch_2012,Veitch_2014,Bravyi_2019,PhysRevLett.123.020401,PRXQuantum.3.020333,Leone_2022,nakahara2008quantum,Leone_2024,bittel2025operationalinterpretationstabilizerentropy,RevModPhys.91.025001}. This formalism provides a clear demarcation between quantum operations that are classically simulable and those that provide a genuine computational advantage. The fundamental building blocks of this analysis are the Pauli strings. For an $n$-qubit system, the phase-free Pauli set $\mathcal{P}_n$ is constructed from tensor products of the standard single-qubit Pauli matrices $\{X,Y,Z\}\equiv\{ \sigma_1, \sigma_2, \sigma_3\}$ and identity matrix $I$. Its elements are explicitly defined as~\cite{Nielsen_Chuang_2010}:
\begin{align}
\mathcal{P}_n = \left\{ \bigotimes_{i=1}^n P_i \ \middle|\ 
\begin{aligned}
&P_i \in \{I, X, Y, Z\}
\end{aligned} \right\}.
\end{align}
The full Pauli group may also include global phases $\phi\in\{\pm1,\pm i\}$, but these phases are physically unobservable and are omitted in the phase-free set used below.
Within this Hilbert space, states that can be efficiently simulated by a classical computer under Clifford evolution are termed stabilizer states. A stabilizer state $\ket{\psi}$ is uniquely determined by an abelian subgroup of $\mathcal{P}_n$ with $2^n$ elements; equivalently, it is the simultaneous eigenstate of all elements in this subgroup. These states form the operational basis of Clifford circuits, which, according to the Gottesman-Knill theorem~\cite{Gottesman_1998,gottesman1998heisenbergrepresentationquantumcomputers}, offer no exponential speedup over classical computation. Universal quantum computation therefore requires nonstabilizer, or Magic, states that cannot be prepared using Clifford gates alone.
To quantify the quantum complexity that cannot be efficiently simulated classically, we utilize the Second Order Stabilizer Rényi Entropy, denoted as $M_2$~\cite{Leone_2022}. Physically, this metric characterizes the extent to which a quantum state is delocalized over the basis formed by the Pauli group elements. It is explicitly constructed from the probability distribution $\Xi_P$, which corresponds to the normalized squared expectation values of Pauli strings for a pure state $\ket{\psi}$:
\begin{align}
    \Xi_P(\ket{\psi}) &= \frac{1}{d} \left|\bra{\psi} P \ket{\psi}\right|^2 ,
\end{align}
where $d = 2^n$ is the dimension of the Hilbert space and $P$ is an element in the phase-free Pauli set. $M_2$ is defined as the Rényi entropy of this distribution, which can be explicitly evaluated as:
\begin{align} \label{eq:M2}
    M_2(\ket{\psi}) &= -\log \left( \sum_{P \in \mathcal{P}_n} \Xi_P^2 (\ket{\psi}) \right) - \log d \nonumber \\
    &= -\log \left( \sum_{P \in \mathcal{P}_n} \frac{1}{d} \left|\bra{\psi} P \ket{\psi}\right|^4 \right).
\end{align}
For a stabilizer state, the fourth moment in Eq.~\eqref{eq:M2} saturates the stabilizer value and gives $M_2=0$. Conversely, a nonzero $M_2$ signals nonstabilizerness, i.e., the Magic resource that obstructs efficient Clifford simulation.
For the specific case of 2-to-2 scattering (a two-qubit system with Hilbert space dimension $d=4$), the spectrum of achievable Magic is rigorously bounded. Notably, two-qubit states obey a constraint strictly tighter than the conjectured general upper bound of $\log\frac{d+1}{2}$. The maximal attainable value is given by $M_2^{\text{max}} = \ln(16/7) \approx 0.827$~\cite{liu2025maximalmagictwoqubitstates}. This theoretical maximum, saturated by states associated with Weyl-Heisenberg group orbits, serves as a fundamental benchmark, enabling us to quantify how close the scattering process comes to saturating the maximal quantum complexity allowed in a two-qubit system.

In the numerical analysis below, Magic generation is defined operationally by applying the spin map to stabilizer inputs and evaluating the Magic of the output state. The number of pure stabilizer states for $n$ qubits is
\begin{equation}
    N_{\rm Stab}(n)=2^n\prod_{j=1}^{n}(2^j+1),
\end{equation}
so for $n=2$ there are $N_{\rm Stab}(2)=60$ input states. Averaging over this finite set gives a basis-independent diagnostic of the non-Clifford content of the conditional two-qubit spin map.

\section{Effective Potential and Spin-Sector Reduction} \label{Sec:potential}
We extend the quantum information framework to low-energy QCD by quantifying the Magic generated in the spin sector of elastic neutron-proton ($np$) scattering. The construction below should be read as a leading spin-effective diagnostic: it retains the OPE spin structures that distinguish CPC and CPV interactions, absorbs channel-dependent normalization factors into effective couplings, and asks how much non-stabilizerness the resulting conditional spin map can generate.

In QCD, the non-trivial topology of the $SU(3)_C$ gauge fields allows a vacuum angle $\theta$. Equivalently, the most general gauge-invariant QCD Lagrangian contains a CP-violating topological term unless the physical angle vanishes~\cite{tHooft:1976rip,Callan:1976je}:
\begin{equation}
    \begin{aligned}
        \mathcal{L}_{\text{QCD}} =& -\frac{1}{4} G_{\mu\nu}^a G^{a\mu\nu} + \sum_f \bar{q}_f (i\gamma^\mu D_\mu - m_f) q_f\\& + \frac{\theta g_s^2}{32\pi^2} G_{\mu\nu}^a \tilde{G}^{a\mu\nu},
    \end{aligned}    
\end{equation}
where $G_{\mu\nu}^a$ is the gluon field strength tensor, $\tilde{G}^{a\mu\nu}$ its dual, and $f$ denotes quark flavors. The parameter $\theta$ serves as a fundamental measure of CPV strength within the strong sector.

However, the Lagrangian parameter $\theta$ is not itself the physical observable~\cite{CREWTHER1979123,Peccei_2008}. The $\theta$ term can be shifted into the quark mass matrix by a chiral rotation~\cite{CREWTHER1979123,Peccei_2008}. If one of the quarks were massless, this phase would become unphysical~\cite{Peccei_2008}. In the Standard Model, the quark Yukawa couplings can also contain CP-violating phases, which contribute to the physical strong CP parameter through the phase of the quark mass matrix. Consequently, the physically observable parameter is the effective combination of $\theta$ and phases of the mass matrix (denoted hereafter as $\bar{\theta}$)~\cite{CREWTHER1979123,Peccei_2008}:
\begin{equation}
    \bar{\theta} = \theta + \text{arg det}(M_q),
\end{equation}
where $M_q$ is the quark mass matrix.

Experimental constraints on $\bar\theta$ are remarkably stringent. Precision measurements of the neutron electric dipole moment (nEDM) impose an effective upper bound of $|\bar{\theta}| \lesssim 10^{-10}$~\cite{Abel_2020}. The unnatural smallness of $\bar{\theta}$, in the absence of any protective symmetry, constitutes the renowned Strong CP Problem. Among the few viable solutions proposed to address this puzzle, the Peccei-Quinn axion mechanism~\cite{Peccei:1977hh,Wilczek:1977pj,PhysRevLett.40.223} stands out as the leading paradigm, providing a simple dynamical explanation for this fine-tuning.

To analyze quantum dynamics at the hadronic scale, we utilize Chiral Perturbation Theory ($\chi$PT) to describe the low-energy Effective Field Theory (EFT) for pion-nucleon interactions~\cite{Weinberg:1968de,Weinberg:1978kz}. We only focus on the first generation sector, so the global symmetry breaking pattern is $SU(2)_L \times SU(2)_R/SU(2)_V$. The corresponding Goldstone bosons form a pion triplet
\begin{equation}
\bm{\tau}\cdot\bm{\pi}
\equiv
\begin{pmatrix}
\pi^0 & \sqrt{2}\pi^+ \\
\sqrt{2}\pi^- & -\pi^0
\end{pmatrix},
\end{equation}
where $\bm{\tau}$ denotes the Pauli matrices acting in isospin space.  
 The proton and neutron are organized into an isospin doublet of $SU(2)_V$, $N=(p,n)^T$. Based on CCWZ construction~\cite{Coleman:1969sm,Callan:1969sn}, the leading CPC pion-nucleon interaction is the axial derivative coupling. For the on-shell non-relativistic reduction used below, we employ the equivalent pseudoscalar form:
\begin{equation}
    \label{eq:L_CPC}
    \mathcal{L}_{\pi NN} = -i g_{\pi NN} \bar{N} \gamma_5 \bm{\tau}\cdot\bm{\pi} N.
\end{equation}
Sub-leading interactions arise from higher-order terms in the chiral expansion, including multi-pion vertices and derivative operators. In the leading low-energy analysis pursued here, we retain the one-pion-exchange contribution because it captures a simple, physically motivated spin structure relevant for the two-qubit description of elastic $np$ scattering. We do not claim that OPE is the complete leading low-energy $NN$ amplitude; contact interactions and rescattering effects are either absorbed into effective coefficients or left to a full EFT analysis beyond the scope of the present work~\cite{Epelbaum_2009, Machleidt:2011zz}.
Therefore, our main numerical results should be understood as applying to the OPE-retained spin-effective description. The persistence of the Magic minimum in a complete chiral EFT scattering treatment is controlled by the corresponding Magic susceptibility around $\bar{\theta}=0$, and requires a systematic analysis beyond the scope of the present work. Similarly, the leading CPV interactions can be obtained from the quark mass spurion terms~\cite{Srednicki:2007qs,CREWTHER1979123},  
\begin{equation}
    \label{eq:L_CPV}
    \mathcal{L}_{CPV}^{\pi N} =
    \bar{g}_{0}\,\bar{N}\bm{\tau}\cdot\bm{\pi}N
    +\bar{g}_{1}\,\pi^0\bar{N}N+\cdots .
\end{equation}
The isoscalar coupling $\bar{g}_{0}$ induced by the QCD $\bar{\theta}$ term is related to the physical vacuum angle through the Crewther-Di Vecchia-Veneziano-Witten (CDVW) relation. Based on numerical estimates from the baryon mass spectrum, it is of order $\bar{g}_{0}\sim 10^{-2}\bar{\theta}$, while $\bar g_1$ is isospin-breaking suppressed for the $\bar\theta$ source~\cite{Pospelov:2005pr,Peccei_2008}.

Focusing on the elastic $n-p$ scattering process, we derive the effective nucleon-nucleon potential by evaluating the tree-level scattering amplitude mediated by single-pion exchange. In the non-relativistic limit, the leading momentum-space one-pion-exchange (OPE) potentials are
\begin{equation}
    \begin{aligned}
        V_{\text{CPC}}(\vec{q}) = - \frac{g_{\pi NN}^2}{4 m_N^2} (\bm{\tau}_1 \cdot \bm{\tau}_2) \frac{(\bm{\sigma}_1 \cdot \bm{q}) (\bm{\sigma}_2 \cdot \bm{q})}{\vec{q}^2 + m_\pi^2}, 
    \end{aligned}
\end{equation}
where $\bm q$ is the three-momentum transfer, $g_{\pi NN}$ is the strong pion-nucleon coupling, and $m_N$ and $m_\pi$ are the nucleon and pion masses. The $s$-wave spin projection of the CPC OPE follows from the angular average
\begin{equation}
    \left\langle q_i q_j F(q^2)\right\rangle_\Omega
    =
    \frac{\delta_{ij}}{3}q^2F(q^2),
\end{equation}
where $\langle\cdots\rangle_\Omega$ denotes averaging over the direction of $\bm q$ at fixed $q^2$,
which turns $(\bm\sigma_1\cdot\bm q)(\bm\sigma_2\cdot\bm q)$ into
$(q^2/3)\bm\sigma_1\cdot\bm\sigma_2$ after angular averaging. The radial and channel-dependent factors are then absorbed into the effective CPC spin phase $f_{\rm CPC}$ introduced below.
\begin{equation}
    \begin{aligned}
        V_{\text{CPV}}(\vec{q})
        &=
        - i \frac{g_{\pi NN}}{2 m_N}
        \frac{\mathcal O_{\rm CPV}(\bm q)}
        {\vec{q}^2 + m_\pi^2},\\
        \mathcal O_{\rm CPV}(\bm q)
        &= \bar g_0(\bm{\tau}_1 \cdot \bm{\tau}_2)
        [(\bm{\sigma}_1 - \bm{\sigma}_2)\cdot\bm q]
        +\bm{\mathcal I}_1\cdot\bm q .
    \end{aligned}
\end{equation}
Here, $\bm{\sigma}_{1,2}$ and $\bm{\tau}_{1,2}$ act on the spin and isospin spaces of the two nucleons, respectively. The $\bm \sigma$ structure is generated from the non-relativistic limit of the nucleon spinors, while the $\tau$ structure comes from the pion matrix in the interaction vertex. The coupling $\bar g_0$ is the leading isoscalar CPV pion-nucleon coupling, and the vector operator $\bm{\mathcal I}_1$ denotes the subleading isospin-breaking structures proportional to the isovector CPV coupling $\bar g_1$ and to higher-order CPV couplings. In a fixed total-isospin channel, $\bm\tau_1\cdot\bm\tau_2$ is replaced by its eigenvalue $C_I=2I(I+1)-3$, namely $C_0=-3$ or $C_1=1$. In the numerical spin model below we retain the dominant $\bar g_0$ channel and absorb the chosen $np$ isospin factor into the effective CPV coefficient.

To build the reduced spin map, we now project these interactions onto the low-energy spin structures retained in the main text. The CPC elastic channel is dominated by the $s$-wave component, so after angular averaging the CPC OPE term reduces to a Heisenberg-type spin-spin coupling~\cite{Epelbaum_2009}. The CPV OPE interaction is parity odd and should be viewed as a perturbation that mixes opposite-parity partial waves rather than as a pure $s$-wave contribution. For a fixed observed momentum-transfer direction, we condition on the scattering kinematics and retain the induced action on the two spin states. This defines a spin map rather than the full orbital-spin scattering channel. Averaging over $\hat n$ in the numerical analysis is therefore a classical average over scattering directions, not a replacement for a full partial-wave EFT calculation. For the full scattering problem, the corresponding spin object at fixed initial momentum and postselected final kinematics would be the operator-valued amplitude
\begin{equation}
    \mathcal S_{\rm spin}^{\rm full}(\hat q)
    =
    \langle \bm p_f(\hat q)|S_{NN}|\bm p_i\rangle ,
\end{equation}
acting on the two-nucleon spin space. Here $S_{NN}$ is the full nucleon-nucleon scattering operator, $\bm p_i$ is the initial relative momentum, and $\bm p_f(\hat q)$ denotes a final relative momentum chosen such that the momentum transfer $\bm q=\bm p_f-\bm p_i$ points along $\hat q$. In the reduced spin ansatz below we write this direction as $\hat n\equiv\hat q$. Given a pure input spin state $\ket{\psi_i}$, the postselected spin output would be
\begin{equation}
    \ket{\psi_f^{\rm full}(\hat q)}
    =
    \frac{\mathcal S_{\rm spin}^{\rm full}(\hat q)\ket{\psi_i}}
    {\left\|
    \mathcal S_{\rm spin}^{\rm full}(\hat q)\ket{\psi_i}
    \right\|},
\end{equation}
and the corresponding Magic diagnostic would be built by averaging $M_2(\ket{\psi_f^{\rm full}(\hat q)})$ over stabilizer inputs and over $\hat n$. In the present work we do not compute $\mathcal S_{\rm spin}^{\rm full}$ from a complete partial-wave chiral EFT amplitude. Instead, we approximate this postselected spin transition operator by a low-energy unitary proxy generated by the leading OPE-inspired spin structures.
With this interpretation, we construct the following spin-effective Hamiltonian:
\begin{equation}\label{Heff}
    H_{\text{eff}}
    =
    f_{\text{CPC}} \, \bm{\sigma}_1 \cdot \bm{\sigma}_2
    + \epsilon \, (\bm{\sigma}_1 - \bm{\sigma}_2) \cdot \hat{n},
\end{equation}
where the first term is the CPC spin-exchange piece and the second term is the direction-dependent CPV perturbation. Here $\hat n\equiv \hat q=\bm q/|\bm q|$ denotes the direction of the momentum transfer, and therefore the direction of the CPV spin perturbation retained in the reduced spin ansatz. The dimensionless coefficients are integrated spin phases: schematically
\begin{equation}
    f_{\rm CPC}\sim \int dt\, C_I V_{\rm CPC}^{\rm spin}(t),
    \qquad
    \epsilon \sim \int dt\, C_I \bar g_0 V_{\rm CPV}^{\rm spin}(t),
\end{equation}
where $t$ denotes the interaction time along the scattering history, and $V_{\rm CPC}^{\rm spin}(t)$ and $V_{\rm CPV}^{\rm spin}(t)$ are schematic spin-projected CPC and CPV potentials. The common kinematic factors, the isospin eigenvalue, and the finite collision time have been absorbed into the effective couplings. Thus $\epsilon$ is linear in $\bar\theta$ in the controlled small-$\bar\theta$ regime, while $f_{\rm CPC}$ is treated as an independent effective CPC spin-exchange phase. The conditional spin map is then represented by
\begin{equation}
    U(\hat n)=e^{-iH_{\rm eff}(\hat n)} .
\end{equation}
This unitary ansatz is not meant to replace the full orbital-spin $S$ matrix; rather, it is a proxy for the postselected spin transition operator $\mathcal S_{\rm spin}^{\rm full}(\hat n)$ in the restricted low-energy spin sector. In the numerical diagnostic below, $U(\hat n)$ is probed by averaging $M_2$ over all 60 pure two-qubit stabilizer inputs and then over scattering directions. The special Clifford point can be seen analytically from the identity

\begin{equation}
    {\rm SWAP}=\frac{1}{2}\left(\mathbf 1+\bm\sigma_1\cdot\bm\sigma_2\right),
\end{equation}
which implies
\begin{equation}
    e^{-i\frac{\pi}{4}\bm\sigma_1\cdot\bm\sigma_2}
    =
    e^{-i\pi/4}\,{\rm SWAP}.
\end{equation}
Thus, when $f_{\text{CPC}} = \pi/4$ and the CPV term vanishes ($\epsilon \to 0$), the operator $U(\hat n)$ reduces to the SWAP gate up to an overall phase, which is a canonical Clifford operation that generates zero Magic. It is useful to clarify how the effective spin-exchange phase
\(f_{\rm CPC}\) can be related to physical low-energy scattering
phases.  In the elastic \(S\)-wave approximation, the CPC
spin part of the physical \(np\) scattering matrix can be written
in the singlet/triplet basis as
\begin{equation}
    S_{\rm phys}^{S{\rm -wave}}
    =
    e^{2i\delta_s(p)}P_s
    +
    e^{2i\delta_t(p)}P_t ,
\end{equation}
where
\begin{equation}
    P_s=\frac{1-\bm\sigma_1\cdot\bm\sigma_2}{4},
    \qquad
    P_t=\frac{3+\bm\sigma_1\cdot\bm\sigma_2}{4},
\end{equation}
and \(\delta_s\equiv\delta_{{}^1S_0}\) and
\(\delta_t\equiv\delta_{{}^3S_1}\) are the physical singlet and
triplet \(S\)-wave phase shifts.  On the other hand, the CPC
part of the spin ansatz gives
\begin{equation}
    e^{-if_{\rm CPC}\bm\sigma_1\cdot\bm\sigma_2}
    =
    e^{3if_{\rm CPC}}P_s
    +
    e^{-if_{\rm CPC}}P_t .
\end{equation}
Since an overall phase does not affect the pure-state Magic
diagnostic used here, the relevant quantity is the singlet-triplet
relative phase.  Matching the relative phases gives
\begin{equation}
    e^{-4if_{\rm CPC}}
    =
    e^{2i[\delta_t(p)-\delta_s(p)]},
\end{equation}
or equivalently
\begin{equation}
    f_{\rm CPC}^{\rm phys}(p)
    =
    -\frac{1}{2}\left[\delta_t(p)-\delta_s(p)\right]
    \quad {\rm mod}\ \frac{\pi}{2}.
\end{equation}
Thus \(f_{\rm CPC}\) should be interpreted as an effective
singlet-triplet relative spin phase, rather than as a microscopic
coupling constant.  In the present work we scan \(f_{\rm CPC}\)
as a phenomenological spin-exchange phase; inserting empirical
or chiral-EFT phase shifts into the relation above gives the
corresponding physical trajectory \(f_{\rm CPC}^{\rm phys}(p)\). The approximation in Eq.~\eqref{Heff} keeps only the dominant spin structures needed for this Clifford point and for the leading CPV deformation. Higher-order kinematic corrections mainly renormalize the effective coefficient multiplying $\bm{\sigma}_1\cdot\bm{\sigma}_2$ after the $s$-wave projection, as discussed in App.~\ref{App:detail}. By contrast, higher partial waves, derivative contact interactions, multi-pion exchange, final-state rescattering, and subleading CPV operators can introduce additional spin-momentum structures. Equation~\eqref{Heff} should therefore be read as a leading low-energy spin-effective description, not as a complete chiral EFT scattering amplitude. Within this controlled regime, the question becomes whether the Magic minimum at $\bar\theta=0$ is stable under variations of the effective CPC spin phase.

\section{Numerical Results and Discussion} \label{Sec:numerical}
\begin{figure}[t]
    \centering
    \includegraphics[width=\columnwidth]{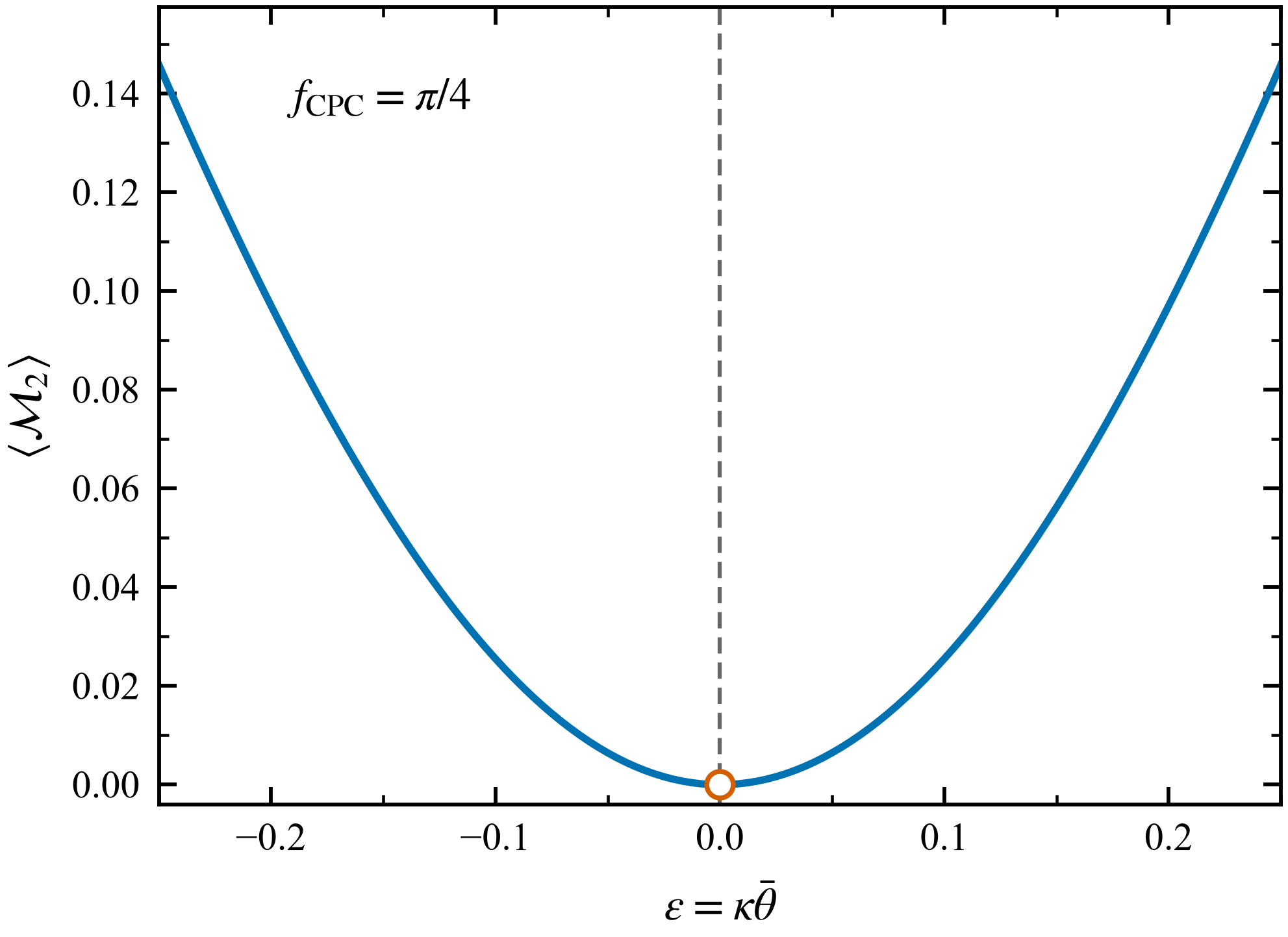}
    \caption{\textbf{The complexity landscape at the CPC Clifford point in the leading spin-effective model.} The solid purple line represents the average Stabilizer Rényi Entropy $\langle \mathcal{M}_2 \rangle$ plotted against the scaled CPV coordinate $\epsilon=\kappa\bar\theta$. The red dot indicates the local minimum at the CP-conserving point $\bar{\theta}=0$, equivalently $\epsilon=0$, within this effective description.}
    \label{fig:magic_min}
\end{figure}

\begin{figure}[t]
    \centering
    \includegraphics[width=\columnwidth]{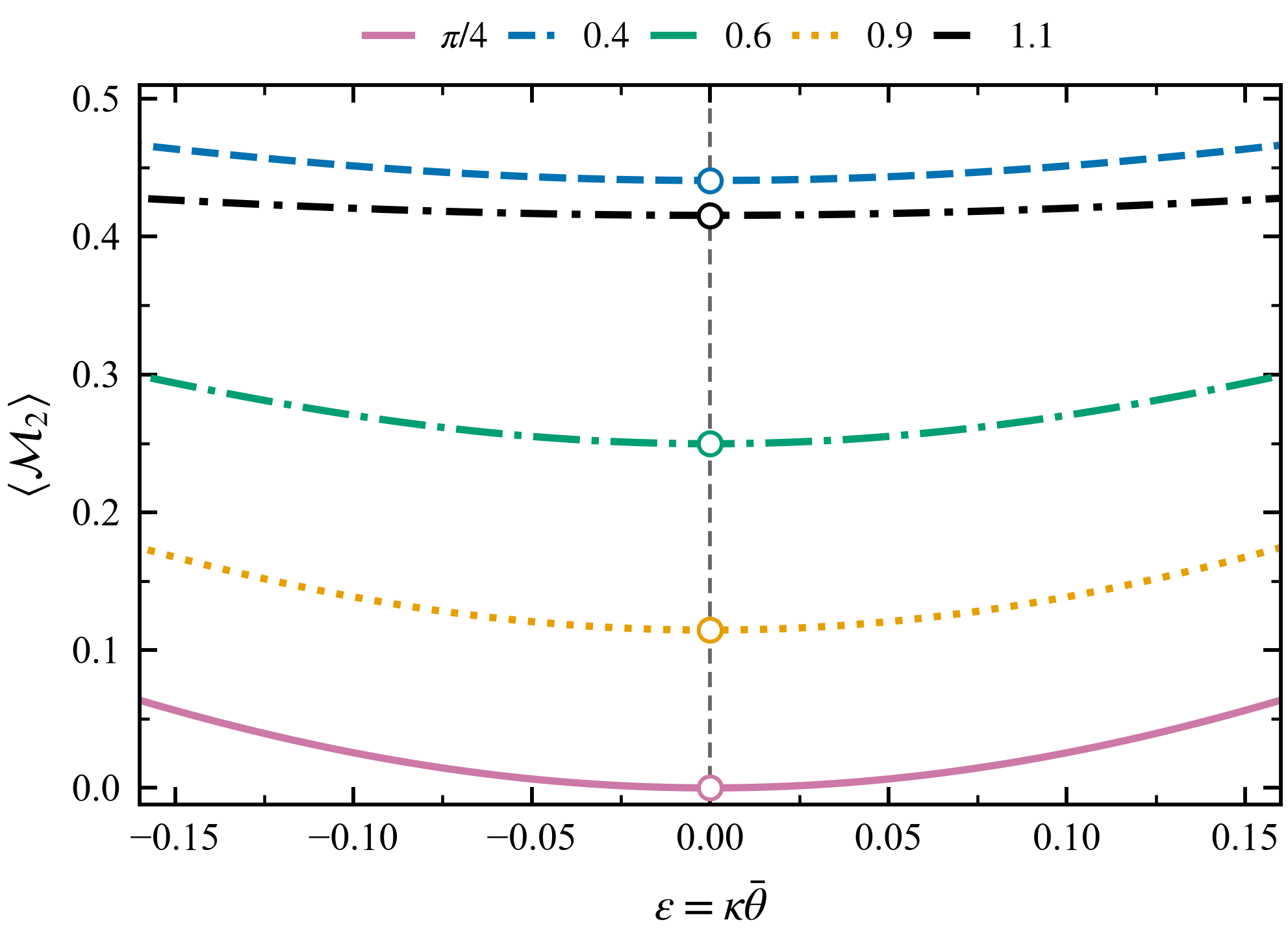}
    \caption{\textbf{Robustness under variations of the effective CPC coupling.} The average Stabilizer Rényi Entropy $\langle \mathcal{M}_2 \rangle$ is plotted against the same scaled CPV coordinate $\epsilon=\kappa\bar\theta$ for several values of the effective CPC phase $f_{\rm CPC}$. While deviations from the Clifford point ($f_{\rm CPC}=\pi/4$, solid purple) lift the baseline complexity due to non-Clifford CPC scattering, the minimum remains located near $\bar{\theta}=0$ for the representative CPC backgrounds shown here. This indicates that the minimum is not solely an artifact of tuning the CPC interaction exactly to the Clifford point.}
    \label{fig:universality}
\end{figure}

This section presents the sampled-point numerical evidence for the local minimum within the unitary spin proxy $U(\hat n)$; the complete finite-dimensional spin-sector functional and its curvature scan are derived in Sec.~\ref{Sec:principle}. We work in the perturbative regime $|\bar{\theta}|\ll 1$, where the leading CPV chiral coupling is linear in $\bar{\theta}$. For a fixed momentum-transfer direction \(\hat n\equiv\hat q\) and interaction parameter \(\bar{\theta}\), the Stabilizer Rényi Entropy \(M_2\) of the final spin state \(\ket{\psi_f(\hat n)}\) was computed from Eq.~\ref{eq:M2}, and averaged over the ensemble of all 60 two-qubit stabilizer initial states $\ket{\psi_i}$:
\begin{align}
 \mathcal{M}_2(\bar{\theta};\hat n)  &=
 \frac{1}{N_{\rm Stab}(2)} \sum_{|\psi_i\rangle \in \mathrm{Stab}_2}
 M_2 \left( \ket{\psi_f(\hat n)}\right) , \nonumber \\
 \ket{\psi_f(\hat n)} &= U(\hat n) \ket{\psi_i} . 
\end{align}
Here $\mathrm{Stab}_2$ denotes the set of all pure two-qubit stabilizer states, so $N_{\rm Stab}(2)=60$.
In the plots, the CPV spin phase is parameterized as
\begin{equation}
    \epsilon=\kappa\,\bar\theta,
\end{equation}
where the dimensionless constant $\kappa$ absorbs the chiral coupling $\bar g_0/\bar\theta$, the fixed isospin factor, kinematic factors, and the integrated collision profile. Rescaling $\kappa$ only rescales the horizontal axis in the perturbative region and does not affect whether $\bar\theta=0$ is a stationary point of the averaged Magic. Accordingly, the horizontal axes in Figs.~\ref{fig:magic_min} and \ref{fig:universality}, as well as in the extended model scan reported in App.~\ref{App:extended_scan}, should be read as a scaled CPV coordinate proportional to $\bar\theta$, rather than as an absolute determination of the physical $\bar\theta$ scale. The numerical question addressed here is therefore the sign and stability of the curvature around $\epsilon=0$ for the spin map in Eq.~\eqref{Heff}.

To characterize the spin-complexity generated without selecting a preferred momentum-transfer direction, we further average $\mathcal M_2$ over $\hat n=\hat q$,
\begin{equation}
    \langle \mathcal{M}_2(\bar{\theta}) \rangle  = \frac{1}{4\pi} \int\mathcal{M}_2(\bar{\theta};\hat n) \sin{\alpha} d \alpha d\beta \,,
\end{equation}
where $\alpha$ and $\beta$ denote the polar and azimuthal angles of the momentum-transfer direction, respectively. This averaging procedure should be interpreted as a uniform directional average of conditional spin maps, independent of specific initial state choices. It is not a cross-section-weighted average of the full $np$ scattering process.

Fig.~\ref{fig:magic_min} gives the first diagnostic at the CPC Clifford point, $f_{\text{CPC}}=\pi/4$. At $\epsilon=0$, the spin map is SWAP up to a global phase, so Clifford evolution maps every stabilizer input to another stabilizer state and the averaged Magic vanishes. Turning on the CPV perturbation moves the output states away from the stabilizer manifold. For the uniform directional average used in the present spin-sector diagnostic, the response is even in $\epsilon$: changing $\epsilon\to-\epsilon$ is equivalent to reversing the momentum-transfer direction $\hat n\to-\hat n$ inside the angular integral. A cross-section-weighted average in the full scattering problem would require the corresponding symmetry analysis of the complete amplitude and is not assumed here. The leading nontrivial response is therefore quadratic, and the numerical curve shows that its coefficient is positive at the Clifford point. Thus the CPC point is a local minimum of the averaged Magic in the leading spin-effective model. This is a statement about the restricted spin-sector diagnostic, not a quantitative derivation of the experimental nEDM bound.

Equivalently, near the CPC point the averaged Magic can be written as
\begin{equation}
    \langle\mathcal M_2\rangle
    =
    \langle\mathcal M_2\rangle_0
    +\frac{1}{2}\chi_M^{\rm eff}\epsilon^2
    +O(\epsilon^4),
\end{equation}
where the absence of a linear term follows from the directional symmetry above. The numerical result in Fig.~\ref{fig:magic_min} corresponds to $\chi_M^{\rm eff}>0$ in the effective spin model.

To demonstrate that this minimum is not an artifact of fine-tuning the interaction strength to the Clifford point, we extended our numerical simulations to a representative range of coupling constants $f_{\text{CPC}}$.

As illustrated in Fig.~\ref{fig:universality}, moving away from $f_{\text{CPC}}=\pi/4$ lifts the CPC baseline because the CPC spin map itself becomes non-Clifford. Nevertheless, for the representative values of $f_{\text{CPC}}$ shown in the figure, the local minimum remains at $\bar{\theta}=0$. This demonstrates that the observed minimum is not solely an artifact of tuning the CPC interaction exactly to the Clifford/SWAP point. Rather, in the leading spin-effective description, the CPV perturbation increases the averaged Magic around the CPC point even when the CPC background already generates nonzero Magic.

An extended scan over a wider range of the model CPV coordinate is shown in App.~\ref{App:extended_scan}. That scan should not be interpreted as a calculation of the full periodic $\theta$ dependence of QCD, because the CPV vertices used here rely on a small-$\bar{\theta}$ expansion~\cite{Srednicki:2007qs}. For the controlled EFT statement, only the neighborhood of $\bar{\theta}=0$ is relevant. Within that neighborhood, the sampled curves show local convexity at the CPC point. Since Figs.~\ref{fig:magic_min} and \ref{fig:universality} probe only representative CPC backgrounds, we next evaluate the finite spin-sector functional itself to determine where this positive curvature persists.

\section{Finite Spin-Sector Magic Functional and the CP-Conserving Minimum}\label{Sec:principle}

The previous section established the local minimum numerically for representative CPC backgrounds. We now evaluate the corresponding finite-dimensional spin-sector Magic functional explicitly, still within the OPE-retained spin ansatz. The object computed in this section is not the full physical $NN$ scattering matrix, but the restricted conditional spin map defined in Eq.~\eqref{Heff}. For each scattering direction, this two-qubit spin map is
\begin{equation}
    \begin{aligned}
    S_{\rm spin}(f_{\rm CPC},\epsilon;\hat n)
    &=
    \exp\Bigl[-i\Bigl(
    f_{\rm CPC}\,\bm\sigma_1\!\cdot\!\bm\sigma_2 \\
    &\qquad\qquad
    +\epsilon\,(\bm\sigma_1-\bm\sigma_2)\!\cdot\!\hat n
    \Bigr)\Bigr].
    \end{aligned}
\end{equation}
Here $f_{\rm CPC}$ is the effective CPC spin-exchange phase introduced in Eq.~\eqref{Heff}, and $\epsilon$ is the corresponding CPV spin phase. For compactness in the finite spin functional below, we write
\begin{equation}
    S_{\hat n}\equiv S_{\rm spin}(f_{\rm CPC},\epsilon;\hat n),
    \qquad \hat n\equiv\hat q .
\end{equation}

\begin{figure*}[t]
    \centering
    \includegraphics[width=0.95\textwidth]{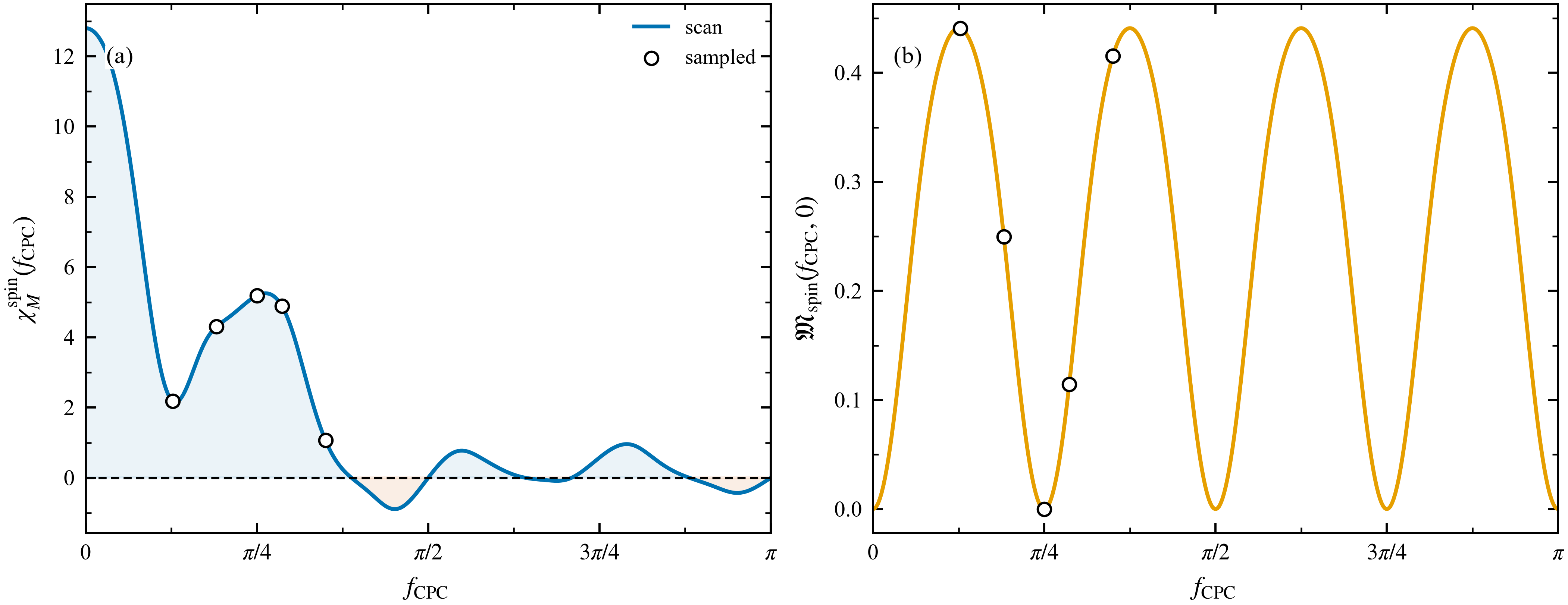}
    \caption{\textbf{Finite spin-sector Magic curvature.}
    The left panel shows a dense scan of the curvature $\chi_M^{\rm spin}(f_{\rm CPC})$ obtained from the complete finite calculation of the restricted spin functional in Eqs.~\eqref{eq:spin_magic_functional}--\eqref{eq:spin_curvature_finite_difference}; the highlighted markers reproduce the sampled CPC backgrounds listed in Table~\ref{tab:spin_susceptibility}. The sign changes outside those sampled points show that positivity is not generic for an arbitrary effective CPC phase. The right panel shows the corresponding CPC baseline Magic $\mathfrak M_{\rm spin}(f_{\rm CPC},0)$.}
    \label{fig:spin_curvature}
\end{figure*}

The corresponding Magic functional is not left implicit. It is computed as
\begin{align}\label{eq:spin_magic_functional}
    \mathfrak M_{\rm spin}(f_{\rm CPC},\epsilon)
    &=
    \frac{1}{4\pi N_{\rm Stab}}
    \int d\Omega_{\hat n}
    \sum_{\psi\in{\rm Stab}_2}
    \mathcal M_\psi(\hat n),
    \nonumber\\
    \mathcal M_\psi(\hat n)
    &=
    M_2\!\left(
    S_{\hat n}\ket{\psi}
    \right),
\end{align}
where $d\Omega_{\hat n}$ is the solid-angle measure for the scattering direction and $N_{\rm Stab}=N_{\rm Stab}(2)=60$. This is the complete finite Magic functional for the restricted spin matrix evaluated in the present work: every pure two-qubit stabilizer input is included, and the scattering direction is averaged over the sphere.

The finite calculation entering Eq.~\eqref{eq:spin_magic_functional} can be written without ambiguity. Let
$A,B\in\mathcal P_2$ be two independent commuting non-identity Pauli strings and
let $s_A,s_B=\pm1$. Each pure two-qubit stabilizer input is the rank-one
projector
\begin{equation}\label{eq:stabilizer_projector}
    \rho_{A,B}^{s_A,s_B}
    =
    \frac{1}{4}
    \left(
    I+s_A A+s_B B+s_A s_B AB
    \right),
\end{equation}
with duplicate projectors removed. This construction generates exactly the 60
distinct pure stabilizer states. For each such input,
\begin{equation}\label{eq:spin_output_density}
    \rho_{\rm out}
    =
    S_{\rm spin}(f_{\rm CPC},\epsilon;\hat n)\,
    \rho_{A,B}^{s_A,s_B}\,
    S_{\rm spin}^{\dagger}(f_{\rm CPC},\epsilon;\hat n),
\end{equation}
and the Pauli moments entering the Stabilizer Rényi entropy are
\begin{equation}\label{eq:pauli_moments}
    r_P(f_{\rm CPC},\epsilon;\hat n)
    =
    {\rm Tr}\!\left(P\rho_{\rm out}\right),
    \qquad P\in\mathcal P_2 .
\end{equation}
Thus the contribution of this input state is evaluated explicitly as
\begin{equation}\label{eq:stabilizer_contribution}
    \mathcal M_{A,B}^{s_A,s_B}(f_{\rm CPC},\epsilon;\hat n)
    =
    -\log\left[
    \frac{1}{4}
    \sum_{P\in\mathcal P_2}
    r_P^4(f_{\rm CPC},\epsilon;\hat n)
    \right].
\end{equation}
Equations~\eqref{eq:spin_magic_functional}--\eqref{eq:stabilizer_contribution}
therefore define a closed finite-dimensional computation:
there are 60 inputs and 16 Pauli moments for each output state, followed only by
the angular average over $\hat n$.

Several features follow directly within the restricted spin functional. Under $\epsilon\to-\epsilon$ together with $\hat n\to-\hat n$, the spin map $S_{\rm spin}(f_{\rm CPC},\epsilon;\hat n)$ is unchanged. Since the angular measure $d\Omega_{\hat n}$ used in Eq.~\eqref{eq:spin_magic_functional} is uniform and invariant under $\hat n\to-\hat n$, the averaged spin-sector functional satisfies
\begin{equation}
    \mathfrak M_{\rm spin}(f_{\rm CPC},\epsilon)
    =
    \mathfrak M_{\rm spin}(f_{\rm CPC},-\epsilon).
\end{equation}
The CP-conserving point is therefore exactly stationary within this uniform-directional spin diagnostic,
\begin{equation}
    \left.
    \frac{\partial\mathfrak M_{\rm spin}}{\partial\epsilon}
    \right|_{\epsilon=0}=0.
\end{equation}
This stationarity should be distinguished from the corresponding statement in a full scattering calculation, where the angular weighting may depend on the physical amplitude and must be analyzed separately. The curvature is then computed directly from this restricted spin matrix,
\begin{equation}
    \chi_M^{\rm spin}(f_{\rm CPC})
    =
    \left.
    \frac{\partial^2\mathfrak M_{\rm spin}(f_{\rm CPC},\epsilon)}
    {\partial\epsilon^2}
    \right|_{\epsilon=0}.
\end{equation}
Operationally, the numbers in Table~\ref{tab:spin_susceptibility} are obtained
from the symmetric second difference
\begin{equation}\label{eq:spin_curvature_finite_difference}
    \begin{aligned}
    \chi_M^{\rm spin}(f_{\rm CPC})
    &=
    \lim_{\delta\to0}
    \frac{\Delta_\delta\mathfrak M_{\rm spin}(f_{\rm CPC})}
    {\delta^2},
    \\
    \Delta_\delta\mathfrak M_{\rm spin}(f_{\rm CPC})
    &=
    \mathfrak M_{\rm spin}(f_{\rm CPC},\delta)
    -2\mathfrak M_{\rm spin}(f_{\rm CPC},0)
    \\
    &\quad
    +\mathfrak M_{\rm spin}(f_{\rm CPC},-\delta).
    \end{aligned}
\end{equation}
In the numerical evaluation we perform the angular average in
Eq.~\eqref{eq:spin_magic_functional} explicitly. This avoids choosing a preferred momentum-transfer direction and defines the curvature as a property of the uniformly direction-averaged spin diagnostic.

\begin{table}[t]
\caption{\label{tab:spin_susceptibility}
Finite spin-sector Magic curvature at $\epsilon=0$ for the sampled CPC backgrounds. The first line is the CPC Clifford/SWAP point.}
\begin{ruledtabular}
\begin{tabular}{c c c}
$f_{\rm CPC}$ & $\mathfrak M_{\rm spin}(f_{\rm CPC},0)$ & $\chi_M^{\rm spin}(f_{\rm CPC})$ \\
\hline
$\pi/4$ & $0$ & $5.18764$ \\
$0.4$ & $0.44068$ & $2.17940$ \\
$0.6$ & $0.24966$ & $4.30374$ \\
$0.9$ & $0.11451$ & $4.89073$ \\
$1.1$ & $0.41537$ & $1.06572$
\end{tabular}
\end{ruledtabular}
\end{table}

Table~\ref{tab:spin_susceptibility} and Fig.~\ref{fig:spin_curvature} show that the finite spin-sector curvature is positive for the Clifford point and for the representative non-Clifford CPC backgrounds explicitly sampled here. The dense scan further resolves three positive-curvature windows in the restricted spin ansatz,
\begin{equation}
    \begin{aligned}\label{eq:positive_curvature_windows}
    0 &\le f_{\rm CPC} \lesssim 1.22,\\
    1.58 &\lesssim f_{\rm CPC} \lesssim 2.02,\\
    2.23 &\lesssim f_{\rm CPC} \lesssim 2.77.
    \end{aligned}
\end{equation}
with zero crossings near $f_{\rm CPC}\approx 1.22$, $\pi/2$, $2.02$, $2.23$, and $2.77$. The representative CPC values used in Figs.~\ref{fig:magic_min} and \ref{fig:universality} all lie inside the first of these positive-curvature windows. Through the phase-shift matching discussed after Eq.~(19), a
physical low-energy $np$ scattering point at momentum $p$
corresponds to a trajectory $f_{\rm CPC}^{\rm phys}(p)$ in this same parameter space. The comparison between this physical trajectory and the positive-curvature windows in Eq.~(\ref{eq:positive_curvature_windows}) provides the direct way to test whether the spin-sector Magic minimum persists for empirical or chiral-EFT phase shifts. We leave a detailed phase-shift numerical implementation of this matching to future work. Since $\epsilon=\kappa\bar\theta$ with real $\kappa$, the curvature with respect to the strong CP angle is
\begin{equation}
    \left.
    \frac{\partial^2\mathfrak M_{\rm spin}}
    {\partial\bar\theta^2}
    \right|_{\bar\theta=0}
    =
    \kappa^2\,\chi_M^{\rm spin}(f_{\rm CPC})
\end{equation}
for these sampled CPC backgrounds. Thus, within the restricted spin matrix defined above and for the CPC phases displayed in Table~\ref{tab:spin_susceptibility}, $\bar\theta=0$ is not merely stationary but a local minimum. The dense scan also shows explicitly that this conclusion is not global in $f_{\rm CPC}$: the sign of $\chi_M^{\rm spin}(f_{\rm CPC})$ reverses for other effective CPC couplings. Extending the same conclusion to the full chiral EFT $S$ matrix would require replacing $S_{\rm spin}$ by the full partial-wave scattering operator and recomputing the corresponding curvature with all low-energy constants included.

\section{Conclusion}\label{Sec:conclusion}
We have shown that, in a leading low-energy spin-effective description of elastic neutron-proton scattering, the CPC point can be characterized as a local minimum of Magic generation. The construction retains the OPE spin structures and treats each scattering direction as a conditional two-qubit spin map. At the CPC Clifford point $f_{\rm CPC}=\pi/4$, the spin map reduces to SWAP up to a global phase and therefore generates zero Magic from stabilizer inputs. For representative non-Clifford CPC backgrounds, the sampled direction-averaged curves remain locally minimized at $\bar{\theta}=0$.

This sampled-point observation was sharpened into an explicit finite spin-sector calculation of the susceptibility,
\begin{equation}
\chi_M^{\rm spin}
=
\left.
\frac{\partial^2\mathfrak M_{\rm spin}}
{\partial\epsilon^2}
\right|_{\epsilon=0}.
\end{equation}
The computed values satisfy $\chi_M^{\rm spin}>0$ for the Clifford point and for the representative non-Clifford CPC backgrounds displayed in Table~\ref{tab:spin_susceptibility}. The dense scan in Fig.~\ref{fig:spin_curvature} further shows that this positivity is confined to specific windows of the effective CPC phase rather than extending to all $f_{\rm CPC}$. Since $\epsilon=\kappa\bar\theta$, the spin-sector Magic is locally minimized at $\bar\theta=0$ precisely in those positive-curvature windows. This gives a concrete information-theoretic sense in which the CP-conserving point is distinguished within the restricted spin sector, echoing similar resource-minimization patterns observed in other contexts~\cite{Liu:2025qfl,Liu:2025bgw}.

The scope of this conclusion is set by the leading low-energy spin-effective description. Higher-order kinematic corrections in the heavy-baryon expansion do not alter the dominant CPC one-pion-exchange spin structure after the $s$-wave projection; they mainly renormalize the effective coefficient of $\bm{\sigma}_1\cdot\bm{\sigma}_2$. However, a complete chiral EFT scattering amplitude contains additional contributions, including higher partial waves, multi-pion exchange, derivative contact operators, final-state rescattering, and possible CPV contact terms. These effects can modify the curvature of the Magic landscape around $\bar{\theta}=0$. The dense scan in Fig.~\ref{fig:spin_curvature} already shows that positivity does not extend to all effective CPC phases even within the restricted spin ansatz. The effective spin phase $f_{\rm CPC}$ can be matched to the physical singlet-triplet $S$-wave phase-shift difference through $f_{\rm CPC}^{\rm phys}(p)
= -[\delta_t(p)-\delta_s(p)]/2$ modulo $\pi/2$, so empirical
phase shifts provide a concrete route for embedding the present
spin-sector windows into physical low-energy \(np\) scattering. Extending the present spin-sector result to the complete chiral EFT S matrix therefore remains a separate partial-wave calculation.

\begin{acknowledgments} 
T.M. is partly supported by Chinese Academy of Sciences Pioneer Initiative "Talent Introduction Plan" (grant No.\,E4ER6601A2), the Fundamental Research Funds for the Central Universities (grant No.\,E4EQ6602X2), and the National Natural Science Foundation of China (grant No.\,E514660101). 
\end{acknowledgments}

\appendix

\section{Derivation of the four nucleon potential} \label{App:detail}
\begingroup

\setlength{\abovedisplayskip}{4pt plus 2pt minus 2pt}

\setlength{\belowdisplayskip}{4pt plus 2pt minus 2pt}

\setlength{\abovedisplayshortskip}{2pt plus 1pt minus 1pt}

\setlength{\belowdisplayshortskip}{4pt plus 2pt minus 2pt}
In the heavy-baryon limit, we perform a non-relativistic reduction of the nucleon bilinears. In the standard Dirac representation, the positive-energy spinor can be written as
\begin{equation}
    u^s(p)
    =
    \sqrt{E_p+m_N}
    \begin{pmatrix}
        \xi^s\\[3pt]
        \dfrac{\bm\sigma\cdot\bm p}{E_p+m_N}\xi^s
    \end{pmatrix},
    E_p=\sqrt{m_N^2+\bm p^2}.
\end{equation}
For $|\bm p|\ll m_N$, this becomes
\begin{equation}
    u^s(p)
    =
    \sqrt{2m_N}
    \begin{pmatrix}
        \left[1+O(\bm p^2/m_N^2)\right]\xi^s\\[3pt]
        \dfrac{\bm\sigma\cdot\bm p}{2m_N}\xi^s
        +O(|\bm p|^3/m_N^3)
    \end{pmatrix}.
\end{equation}
Here $\xi^s$ is a two-component Pauli spinor with spin label $s$. The leading spin dependence follows from the lower component and from the identity
\begin{equation}
    (\bm\sigma\cdot\bm a)(\bm\sigma\cdot\bm b)
    =
    \bm a\cdot\bm b
    +
    i\bm\sigma\cdot(\bm a\times\bm b).
\end{equation}
Consequently, for the elastic kinematics relevant to the leading OPE amplitude, the dominant CPC pseudoscalar vertex remains proportional to $\bm{\sigma}\cdot\bm q$ up to momentum-suppressed coefficient corrections. Spin-orbit structures may appear beyond the leading projection, but they are not part of the $s$-wave spin-spin operator retained in the main text. In the non-relativistic normalization convention appropriate for the potential matching used below, the CPC vertex takes the form
\begin{equation}
    \begin{aligned}
        \mathcal{V}_{\text{CPC}}^{\rm NR}
        &=
        \xi^\dagger
        \frac{\bm{q}\cdot\bm\sigma}{2m_N}
        \left[
        1
        +O\!\left(
        \frac{P^2}{m_N^2},
        \frac{q^2}{m_N^2}
        \right)
        \right]
        \xi \,.
    \end{aligned}
\end{equation}
where $\bm{P} = \frac{1}{2}\left (\bm{p}_1 +\bm{p}_2\right )$ and $\bm{q} = \bm{p}_1 -\bm{p}_2$.
Similarly, after absorbing the overall factor $2m_N$ associated with relativistically normalized external spinors into the non-relativistic matching, the CPV scalar bilinear can be written as 
\begin{equation}
    \begin{aligned}
        \mathcal{V}_{\text{CPV}}^{\rm NR}
        &=
        \xi^\dagger \bigg[
        1
        +O\!\left(
        \frac{P^2}{m_N^2},
        \frac{q^2}{m_N^2}
        \right)
        + \frac{i}{4m_N^2}(\bm{P}\times\bm{q})\cdot\bm\sigma
        \bigg]\xi .
    \end{aligned}
\end{equation}
The second term in $\mathcal{V}_{\rm CPV}^{\rm NR}$ generates momentum-suppressed spin-orbit-type structures. Since these terms are suppressed in the low-energy regime and are proportional to the CPV coupling, we retain only the leading CPV interaction in the main analysis. Such subleading CPV structures can modify the curvature of the Magic landscape but do not shift the CPC point itself.

The CPC potential arises from the exchange of a pion between two CPC vertices. Up to the overall sign convention relating the non-relativistic amplitude and the potential, which is absorbed into the effective spin phase in the main text, the scattering amplitude is constructed as:
\begin{equation}
    \begin{aligned}
    \mathcal M_{\rm CPC}
    &=
    \left[-\frac{g_{\pi NN}}{2m_N}
    (\bm{\sigma}_1\cdot\bm{q})\right]
    \frac{\bm{\tau}_1\cdot\bm{\tau}_2}{q^2+m_{\pi}^2}
    \left[-\frac{g_{\pi NN}}{2m_N}
    (\bm{\sigma}_2\cdot(-\bm{q}))\right].
    \end{aligned}
\end{equation}
Simplifying the expression and extracting the potential with the above convention, we obtain the tensor-like interaction,
\begin{equation}
    \begin{aligned}
        V_{CPC}(\bm q)
        &=-\frac{g_{\pi NN}^2}{4m_N^2}
        (\bm\tau_1\cdot\bm\tau_2)
        \frac{(\bm\sigma_1\cdot\bm q)(\bm\sigma_2\cdot\bm q)}
        {q^2+m_\pi^2}
    \end{aligned}\,.
\end{equation}
The leading-order CPV potential arises from the interference between a CPC (pseudoscalar) vertex and a CPV (scalar) vertex. Keeping the dominant isoscalar CPV coupling $\bar g_0$ for the $\bar\theta$ source, the total scattering amplitude includes two diagrams, again up to the same overall potential convention:
\begin{equation}
    \begin{aligned}
        V_{CPV}^{(0)}(\bm{q})
        &=-\frac{\bm\tau_1\cdot\bm\tau_2}{q^2+m_\pi^2}
        \bigg (
        \left[i\bar g_0\bm{1} \right]_1
        \left [ -\frac{g_{\pi NN}}{2m_N}
        (\bm{\sigma_2}\cdot(\bm{-q}))\right]_2 \\
        &\quad+
        \left [ -\frac{g_{\pi NN}}{2m_N}
        (\bm{\sigma_1}\cdot\bm{q})\right]_1
        \left[i\bar g_0\bm{1} \right]_2
        \bigg) \,.
    \end{aligned}
\end{equation}
Combining these terms leads to the final expression featuring the characteristic spin difference:
\begin{equation}
    V_{CPV}^{(0)}(\bm{q})=
    -i\frac{\bar{g}_{0}g_{\pi NN}}{2m_N}
    \frac{(\bm{\tau}_1\cdot\bm\tau_2)}{q^2+m_\pi^2}
    \left[(\bm{\sigma}_1-\bm{\sigma}_2)\cdot\bm{q}\right]\,.
\end{equation}
Additional isospin-breaking CPV couplings, such as $\bar g_1$, generate further spin-isospin operators. These terms are omitted from the minimal spin Hamiltonian used in the numerical scan and should be included in a full chiral EFT susceptibility analysis.

\endgroup
\vspace{0.6\baselineskip}
\section{Extended model scan} \label{App:extended_scan}

For completeness, we also examined the behavior of the spin-unitary ansatz over a wider range of the model CPV coordinate. Because the CPV coupling used in the main text is derived from a small-$\bar\theta$ expansion, this scan should be interpreted only as a model-level extrapolation of the restricted spin dynamics, not as the full periodic $\theta$ dependence of QCD. Its purpose is to illustrate the global shape of the finite spin ansatz and to distinguish the controlled local susceptibility statement from large-\(\epsilon\) model artifacts.

For the Clifford background $f_{\rm CPC}=\pi/4$, the extended profile remains exactly even in $\epsilon$, so its derivative is odd. In the displayed numerical scan, the derivative vanishes at
\begin{equation}
    \epsilon \approx -0.80,\qquad 0,\qquad 0.80,
\end{equation}
corresponding to one central minimum and two symmetric outer turning points. The outer extrema occur at
\begin{equation}
    \langle \mathcal M_2\rangle_{\rm max}\approx 0.585
    \qquad
    {\rm at}
    \qquad
    |\epsilon|\approx 0.80 .
\end{equation}
These shoulders are useful for visualizing the global shape of the extrapolated spin-unitary ansatz, but they should not be overinterpreted as physical features of the full strong-CP landscape. In particular, they are stationary points of the restricted two-qubit spin proxy, not predictions for the true QCD vacuum-angle dependence.

The controlled EFT statement remains only the local one near $\epsilon=0$: within the perturbative neighborhood where $\epsilon\propto\bar\theta$ is meaningful, the curve is smooth, symmetric, and convex upward at the origin. The additional extrema at $|\epsilon|=O(1)$ lie outside that controlled small-$\bar\theta$ regime and simply indicate where the restricted spin model ceases to resemble a local susceptibility expansion of the full hadronic problem.

\balance
\bibliography{apssamp}

\end{document}